\documentclass{llncs}
\usepackage{amssymb}
\usepackage{amsmath}
\usepackage{graphicx}
\usepackage[font=scriptsize,labelformat=simple]{subfig}

\usepackage{xcolor}
\usepackage{enumitem}
\usepackage{listings}
\usepackage{url}
\usepackage{multirow}
\usepackage{textcomp}
\usepackage{fancyhdr}
\usepackage[author={Mustafa}]{pdfcomment}
\usepackage[utf8]{inputenc}
\usepackage[english]{babel}
\usepackage{mathtools}
\DeclarePairedDelimiter{\ceil}{\lceil}{\rceil}

\usepackage{amsthm}
\theoremstyle{definition}

\usepackage[linesnumbered,ruled]{algorithm2e}

\newcommand{\sbucketsort}{{\tt bucket-sort}}
\newcommand{\salltoall}{{\tt alltoall}}
\newcommand{\sallgather}{{\tt allgather}}
\newcommand{\smpirecv}{{\tt MPI\_Recv}}

\newcommand{\loggp}{{\tt LogGP}}

\newcommand{\smpialltoallv}{{\tt MPI\_Alltoallv}}

\newcommand{\smpialltoallvn}{{\tt MPI\_Neighbor\_alltoallv}}
\newcommand*\xor{\mathbin{\oplus}}
\newcommand{\sstapl}{{\tt STAPL}}
\newcommand{\sfmm}{{\tt FMM}}

\newcommand{\scharm}{{\tt Charm$++$}}

\newcommand{\smtol}{{\tt M2L}}
\newcommand{\sptop}{{\tt P2P}}

\newcommand{\smtom}{{\tt M2M}}
\newcommand{\sptom}{{\tt P2M}}

\newcommand{\smpi}{{MPI}}

\newcommand{\sjugene}{{\tt JUGENE}}
\newcommand{\sexafmm}{{\tt exaFMM}}
\newcommand{\spvfmm}{{\tt PVFMM}}

\begin{document}

%

\title{Communication Reducing Algorithms for Distributed Hierarchical N-Body Problems with Boundary Distributions}

\author{Mustafa Abduljabbar\inst{1} \and George Markomanolis\inst{1} \and Huda Ibeid \inst{1} \and Rio Yokota\inst{2} \and David Keyes\inst{1}}

\authorrunning{Abduljabbar et al.}

\institute{King Abdullah University of Science and Technology,
  Thuwal, Saudi Arabia
  \and
  Tokyo Institute of Technology, Tokyo, Japan
}

\maketitle
\begin{abstract}
Reduction of communication and efficient partitioning are key issues for achieving scalability in hierarchical $N$-Body algorithms like \sfmm. In the present work, we propose four independent strategies to improve partitioning and reduce communication. First of all, we show that the conventional wisdom of using space-filling curve partitioning may not work well for boundary integral problems, which constitute about $50\%$ of \sfmm’s application user base. We propose an alternative method which modifies orthogonal recursive bisection to solve the cell-partition misalignment that has kept it from scaling previously. Secondly, we optimize the granularity of communication to find the optimal balance between a bulk-synchronous collective communication of the local essential tree and an RDMA per task per cell. Finally, we take the dynamic sparse data exchange proposed by Hoefler et al. \cite{hoefler2010} and extend it to a \textit{hierarchical} sparse data exchange, which is demonstrated at scale to be faster than the MPI library's \smpialltoallv\ that is commonly used.
\end{abstract}


\section{Introduction}\label{sec:intro}
The $N$-body problem is a kernel in many scientific simulations where the behavior of the system is defined from mutual interactions between discrete entities (e.g., molecules, charges, astrophysical bodies). The $N$-body algorithm sums up contributions due to all particles in the system and results in quadratic complexity. The Barnes-Hut treecode, which subdivides the 2D/3D domain into quad/octrees, brings the complexity down to $\mathcal{O}(N\log{N})$ by hierarchically clustering the sources into multipole expansions. The fast multipole method (\sfmm) clusters the targets into local expansions to bring the complexity further down to $\mathcal{O}(N)$. For mathematical foundations of the multipole expansions, see \cite{appel1985}, \cite{greengard1987}, and \cite{beatson1997}. Among the applications of \sfmm\ are \cite{lu2006} and \cite{yokota2011} where protein-protein encounter within a biomolecular dynamics solver is accelerated by using \sfmm\ to solve the boundary integral equation, which is used to discretize the linearized Possion-Boltzmann equation. In \cite{ohno2014} all-atom molecular dynamics is performed to simulate the conditions of living cells by calculating energy at target proteins in a solvent and a molecular crowder using \sfmm\/. It is also used to speedup the matrix-vector multiplication, which arises from electromagnetic scattering problems \cite{rui2007}. Other applications include gravity simulations \cite{bedorf2012,price2007energy}.

Due to its increased importance in large-scale simulations, there is now a considerable literature on implementing parallel hierarchical $N$-body solvers. Also, since \sfmm\ is among Berkeley's seven dwarfs, the numerical methods that are believed to be the most impactful in science and engineering according to \cite{asanovic2006landscape}, it is important to address issues arising at exascale especially the increasing cost of data movement (through memory hierarchy or network) as opposed to floating point operations. Even though many of the current \sfmm\ implementations are scalable to the full machine they run on, a communication reducing approach that works on at least an order of magnitude more nodes tends to be rarely the emphasis of these implementations. This tendency is justified in accordance to the trend in enhancing a node with multi/many-core capabilities. However, even within a many-core node, more sophisticated methods should be used to place and exchange data to get the maximum performance reported by the vendor. This is already implied in equipping the second generation of Intel\textsuperscript{\textregistered} Xeon Phi\texttrademark\/ processors code-named Knights Landing (KNL) with memory `clustering modes'. Therefore, ideas presented in this paper complement the literature although they mainly target distributed memory.

An example work that achieves full machine scalability using GPUs is that of B{\'e}dorf et al. \cite{bedorf2012}, where a parallel algorithm for sparse tree construction and traversal that works completely on the GPU is introduced. At the construction phase, they map the 3D coordinates to Hilbert's linear (n-bit) addresses, then particles are sorted to achieve locality in memory. To avoid the typical sequential insertions to build Morton trees \cite{warren1994}, one particle is assigned per GPU thread. A top-bottom mask is applied successively on each particle to discover its predecessors such that cells with less than $N_{leaf}$ are considered leaves. Grouping of particles is done using parallel compact algorithm. To exploit the massively parallel GPU threads, a breadth-first traversal is used to carry out the computation. They report a processing rate of 2.8 million particles per second. This work was extended to an MPI parallel version where they achieved 24.77 PFlop/s (mixed precision) on the full node of Titan \cite{bedorf2014}.

Speck et al. \cite{speck2012} report scalability on up to 262,144 cores by introducing temporal parallelism (parallel-in-time algorithm) on top of MPI/Pthreads spatial decomposition to overcome the strong scaling limits when the number of particles per node becomes too small. The scalability is shown for up to 4M particles; then when they take advantage of shared and distributed memory parallelism, and exploit the overlap of data-exchange and computation, they calculate 2 billion particles on 262,144 cores of \sjugene\ , according to \cite{Winkel2012880}. Lashuk et al. \cite{lashuk2012} propose an \sfmm\ implementation that scales on up to 196,608 cores by providing a novel domain-specific bulk synchronous all-reduce algorithm for remote tree communication. They report communication complexity of $O(\sqrt{P}*(\frac{N}{P})^{2/3})$ , which comes from their hypercube \salltoall\ communication scheme.
Hoefler et al. \cite{hoefler2010} discuss the time and memory complexity of the common protocols used for the dynamic sparse data exchange problem and develop the non-blocking exchange protocol ($\mathcal{NBX}$) with constant memory overhead. Their novel algorithm improves the runtime of sparse data-exchange up to 8,192 processors of Bluegene/P by a factor of 5.6. They prove and model a generic time complexity of O($\log{P}$) using the \loggp\ model.  

Zandifar et al. \cite{zandifar2015} provide a parallel \sfmm\ implementation as a benchmark for their high-level skeletons (abstract parallel patterns) framework which executes on top of the \sstapl\ runtime system that dynamically schedules task on highly heterogeneous architectures. They reuse several parallel patterns like the \sbucketsort\ and \salltoall\ to perform geometric bisection and to aggregate the local essential tree (LET) respectively.  They achieve comparable performance to the corresponding base MPI implementation by taking advantage of the underlying data-driven execution and asynchronous task scheduling guaranteed by the runtime system. Many features of \scharm\ like task migration and Structured Control Flow are augmented in \cite{abduljabbar2014} to overlap computation with the communication of the local essential tree (LET).

Contributions of the present work can be summarized as follows:
\begin{itemize}
  \item A novel demonstration that shows a weakness in Hilbert's space-filling interval partitioning for boundary element distributions.
  \item A communication scheme with adjustable granularity, which enables the overlap of local essential tree communication with computation that otherwise cannot be overlapped.
  \item Introduction of the adaptive hierarchical sparse data exchange ($\mathcal{HSDX}$), a neighborhood collective communication algorithm for exchanging the global tree in a few steps by direct near-field communication only.
\end{itemize}
In section \ref{sec:partition}, we describe our adopted partitioning techniques and justify our choice in detail. Section \ref{sec:communication_methodologies} describes different communication strategies that we adopt in order to avoid bulk synchronous LET communication. We also describe the adopted load-balancing strategies and the communication complexity analysis of our approach. Finally, we demonstrate our scalability and evaluation results.

\section{Partitioning Schemes for FMM}\label{sec:partition}
There are two traditional objectives associated with good partitioning of the $N$-body problem: evenly splitting data among partitions to achieve work balance, and providing efficient access to non-local data. There is no optimal approach that can simultaneously handle these two objectives, because of strict considerations on locality of data for high arithmetic intensity, granularity, and the size of communication which can vary based on space-time proximity of partitions.

\begin{figure}[t]
\centering
\subfloat[HOT (Morton)]{
\includegraphics[width=0.2\textwidth]{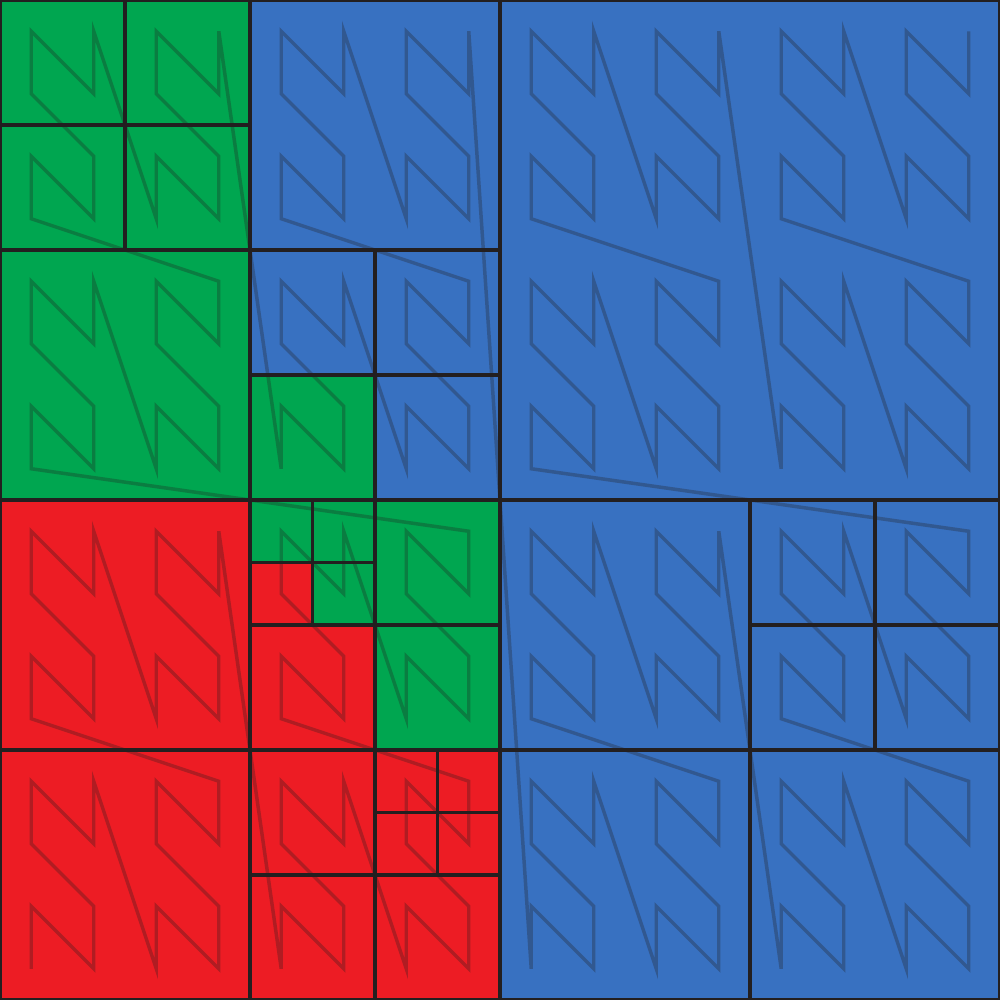}\label{fig:morton}}
\subfloat[HOT (Hilbert)]{
\includegraphics[width=0.2\textwidth]{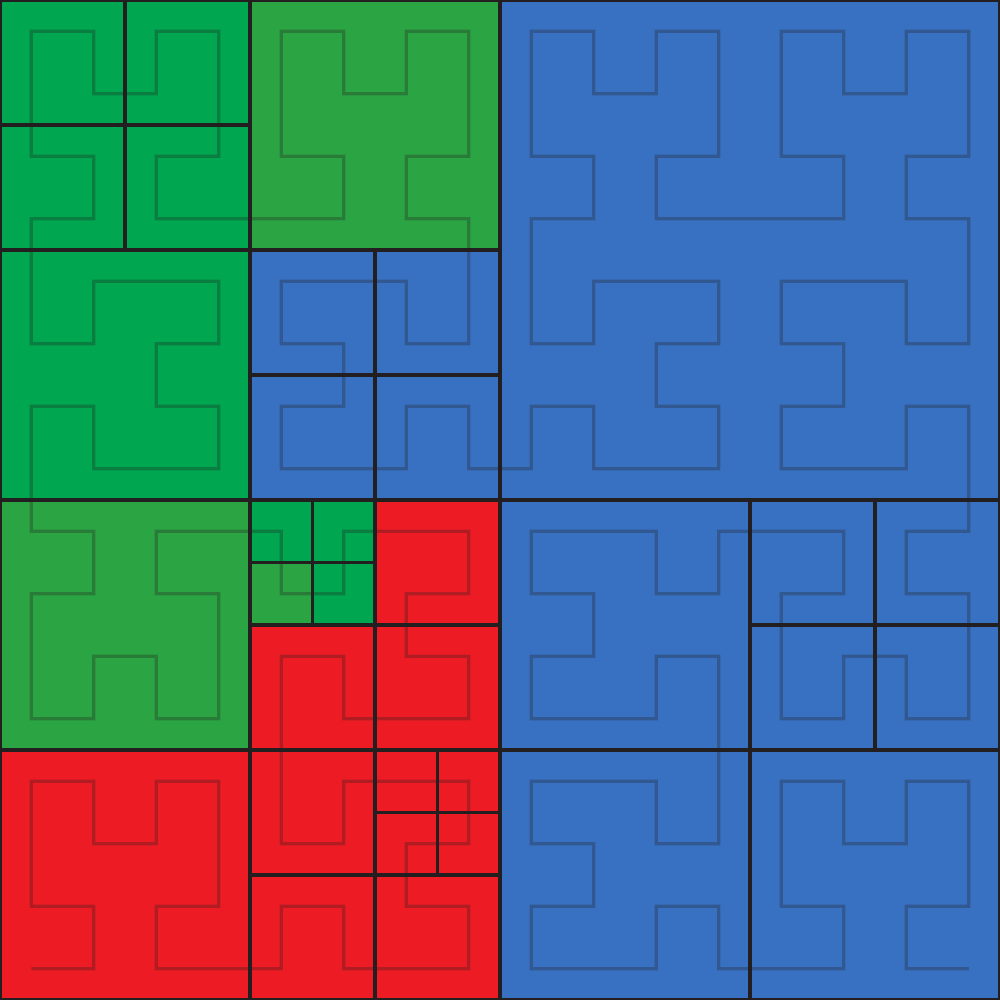}\label{fig:hilbert}}
\subfloat[ORB]{
\includegraphics[width=0.2\textwidth]{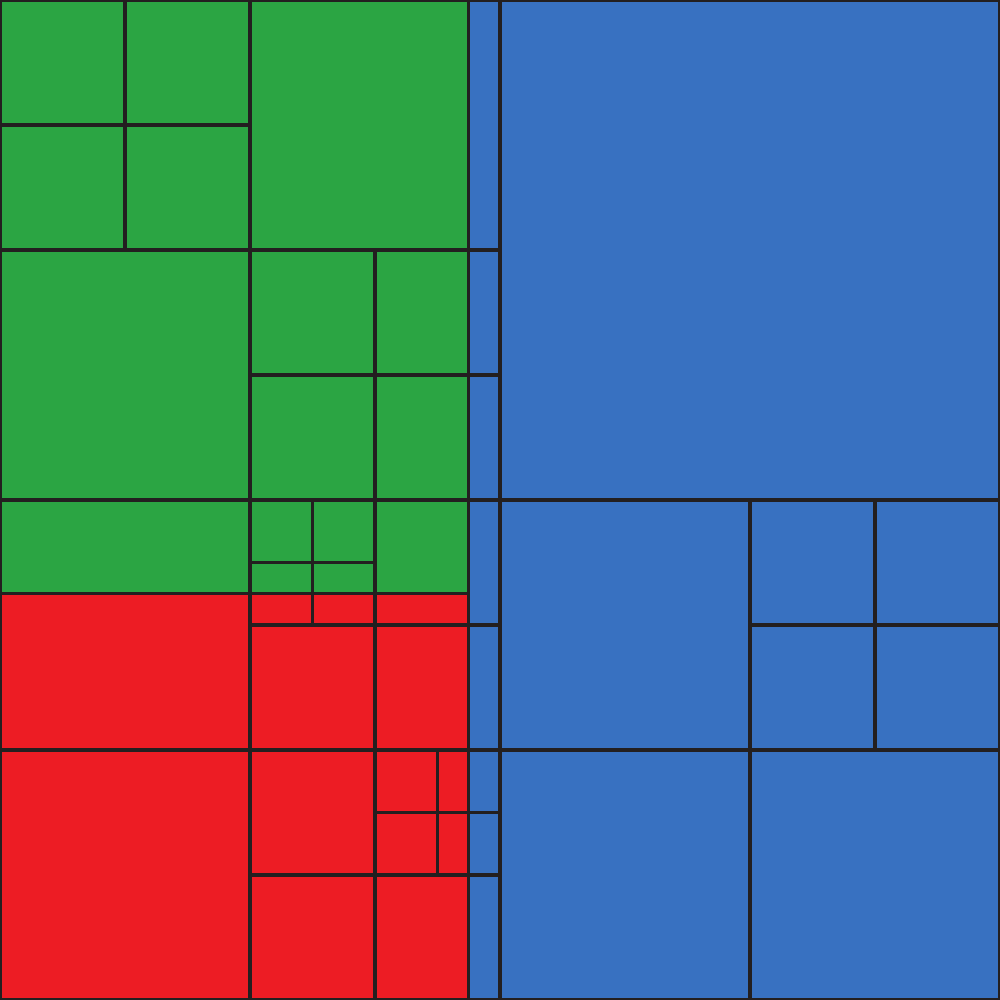}\label{fig:orb}}
\subfloat[Present method]{
\includegraphics[width=0.2\textwidth]{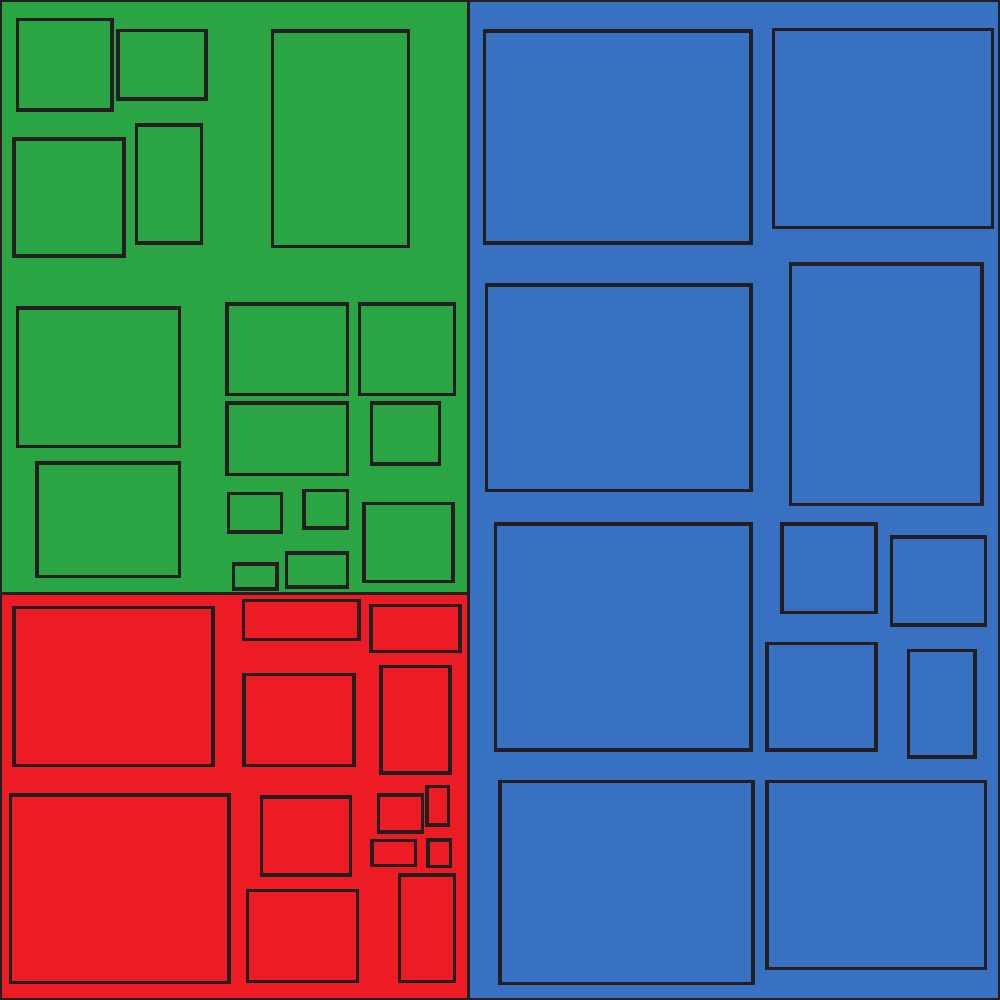}\label{fig:orb2}}\\
\caption{Schematic of different partitioning schemes. (a) shows the hashed octree with Morton keys. (b) shows the hashed octree with Hilbert keys. (c) shows the orthogonal recursive bisection with an underlying global tree. (d) is the present method using an orthogonal recursive bisection with independent local trees and tight bounding boxes.}
\label{fig:partition}
\end{figure}
\subsection{Preliminaries}
Partitioning schemes for fast $N$-body methods can be categorized into orthogonal recursive bisections (ORB) \cite{salmon1991} or hashed octrees (HOT) \cite{warren1993}. 
\subsubsection{Orthogonal Recursive Bisection (ORB)}
The ORB \cite{salmon1991} forms a balanced binary tree by finding a geometric bisector that splits the number of particles equally at every bisection of the tree. The direction of the geometric bisector alternates orthogonally ($x$, $y$, $z$, $x$, ...) to form a cascade of rectangular subdomains that contain equal number of particles similar to Fig~\ref{fig:orb}. For nonuniform distributions the aspect ratio of the subdomain could become large, which leads to suboptimal interaction list size and communication load. This problem can be solved by choosing the direction of the geometric bisector to always split in the longest dimension. The original method is limited to cases where the number of processes is a power of two, but the method can be extended to non-powers-of-two by using multi sections instead of bisections \cite{makino2004}.

\subsubsection{Hashed Oct-Tree (HOT)}
In HOT, initially proposed by \cite{warren1993}, domain is partitioned by splitting Morton/Hilbert ordered space filling curves into equal segments as shown in figures \ref{fig:morton} and \ref{fig:hilbert}. Morton/Hilbert ordering maps the geometrical location of each particle to a single key. The value of the key depends on the depth of the tree at which the space filling curve is drawn. Three bits of the key are used to indicate which octant the particle belongs to at every level of the octree. Therefore, a 32-bit unsigned integer can represent a tree with 10 levels, and a 64-bit unsigned integer can represent a tree with 21 levels. Directly mapping this key to the memory address is inefficient for non-uniform distributions since most of the keys will not be used. Therefore, a hashing function is used to map the Morton/Hilbert key to the memory address of particles/cells.

\begin{figure}[t]
\centering
\includegraphics[width=0.5\textwidth]{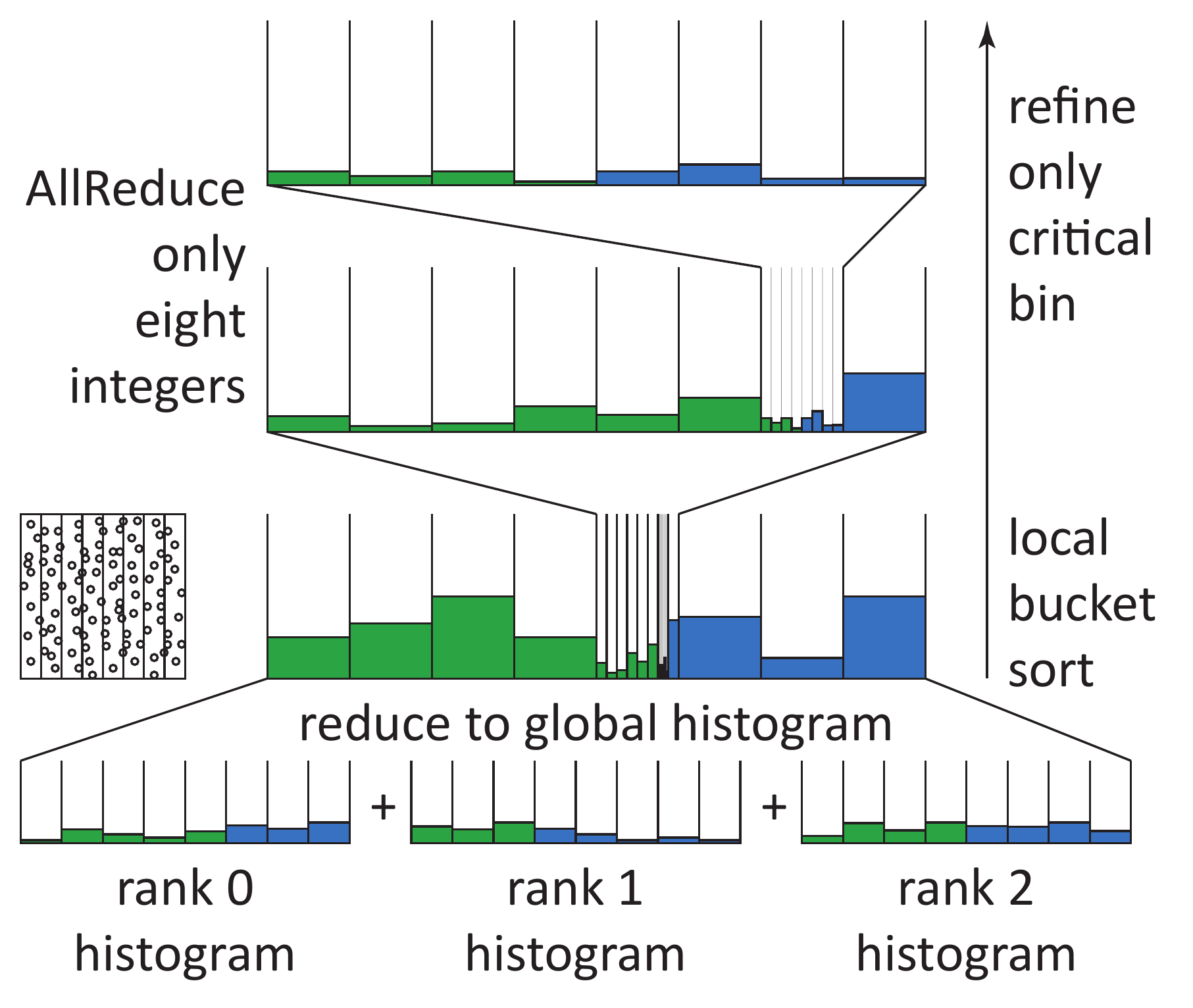}
\caption{Histogram-based partitioning scheme.}
\label{fig:split}
\end{figure}

\subsection{Adopted Partitioning Strategies}
\subsubsection{Parallel Sampling-Based Techniques for Finding Splitters/Bisectors} 
Parallel sampling-based techniques have proven to be useful for both finding the bisectors in ORB \cite{makino2004} and finding the splitting keys in HOT \cite{Solomonik2010}. Both ORB and HOT are constructing parallel tree structures, but in different ways. There is an analogy between parallel tree construction and parallel sorting. The idea behind ORB is analogous to merge sort, where a divide and conquer approach is taken. HOT is analogous to radix sort, where each bit of the key is examined at each step. Therefore, sampling-based techniques that are known to be effective for parallel sorting are also effective for parallel tree partitioning. The partitioning can be separated into two steps. The first step is to find the bisectors/key-splitters by using a sampling-based parallel sorting algorithm. An example of such sampling-based partitioning is shown in Fig~\ref{fig:split}. Sorting is only performed among the buckets (not within them) and this is done only locally. The only global information that is communicated is the histogram counts, which is only a few integers and can be done efficiently with an \texttt{MPI\_allreduce} operation. The bins can be iteratively refined to narrow the search for the splitter of the HOT key or ORB bisector. This will determine the destination process for each particle. The second step is to perform an all-to-all communication of the particles. Since the ORB bisector is one floating point number and the HOT key is one integer, it is much less data than sending around particle data at each step of the parallel sort.

\subsubsection{Weakness in Space-Filling Partitioning for Boundary Distributions} \label{sec:avoiding}
\begin{figure}[t]
     \centering
     \subfloat[Test][Hilbert partition A.]
       {\includegraphics[width=0.33\columnwidth]{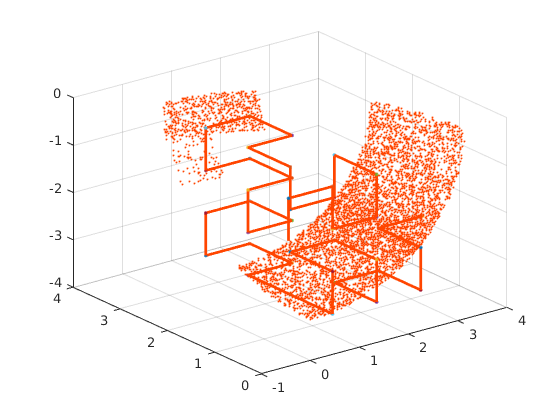}\label{fig:hilbert_problem1}}
     \subfloat[Test][A, B combined.]
     {\includegraphics[width=0.33\columnwidth]{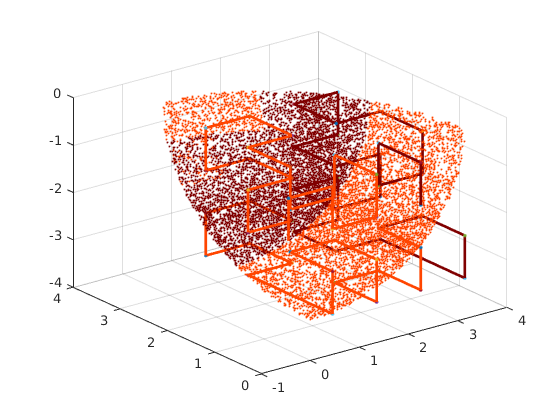}\label{fig:hilbert_problem3}}
     \subfloat[Test][Underlying Hilbert curve.]
     {\includegraphics[width=0.33\columnwidth]{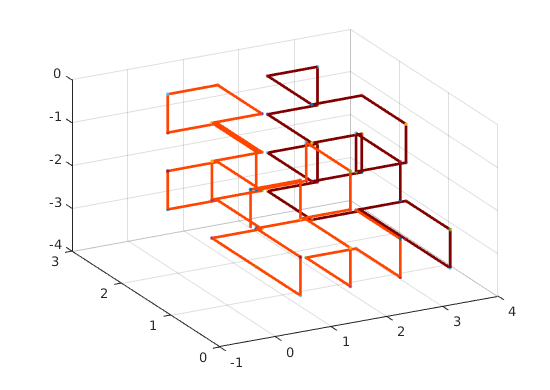}\label{fig:hilbert_problem4}}     

\caption{A Hilbert partition from a boundary spherical distribution viewed from different perspectives. A space discontinuity exists even though partitions are in correct Hilbert order due to the existence of hollow space in orthogonal dimensions.}\label{fig:hilbert_problem}
\end{figure}

It is well-known that the main advantage of Hilbert curve as opposed to Morton is its locality preserving properties in 2D. It is not clear, however, to what extent we can generalize this property in higher dimensions \cite{haverkort2011inventory}. As a counterexample to the locality property, we observe that it is not entirely preserved in case of 3D boundary element distributions, which increases the distributed interaction list size. The reason for that comes from the intuitive notion of space-filling curves, that is, when the space is not filled, e.g., in boundary spherical distribution, interpolation of spatial points to Hilbert curve does not necessarily map to keys that are continuous in space. This is attributable to the fact that keys are not interpolated in their natural order, since point are spread out on surface units. Fig~\ref{fig:hilbert_problem} shows particles laid out in their respective Hilbert order. Due to the geometry of the space-filling curve, movement across dimensions happens orthogonally, hence, if hollow space is encountered in the orthogonal direction, it will introduce discontinuity in the partition as in Fig~\ref{fig:hilbert_problem1}. Clearly, this does not apply to uniform dense distributions, which comprise many classical applications of \sfmm\/, making HOT partitioning an optimal choice in such cases.

\subsubsection{Hybrid Partitioning}\label{sec:hybrid}

In our implementation, we choose modified version of ORB over HOT for a few other reasons. One of the main reasons is that we were able to improve a major defect of ORB -- partition-cell alignment issue. Since geometrically closer points interact more densely with each other, it is crucial to keep the particles in the same cell on the same process in order to minimize communication. However, if a global Morton/Hilbert key is used to construct the local trees, the ORB may place a bisector in the middle of a cell as shown in Figure~\ref{fig:orb}. This results in an increase in the interaction list size. We avoid this problem by using local Morton/Hilbert keys that use the bounds of the local partition. This may at first seem to increase the interaction list near the partition boundaries since two misaligned tree structures are formed. However, when one considers the fact that the present method squeezes the bounding box of each cell to tightly fit the particles as shown in Figure~\ref{fig:orb2}, it can be seen that the cells are not aligned at all in the first place. Furthermore, our flexible definition of the multipole acceptance criteria optimizes the interaction list length for a given accuracy regardless of the misalignment.

\section{Communication of the Local Essential Tree}
Once particles are partitioned, those in the local domain are used to construct a local tree. We use a completely local construction of the octree using the local bounding box, instead of using a global Morton/Hilbert key that is derived from the global bounding box. This allows us to reuse all parts of the serial code and only add a few routines for the partitioning, grafting of trees, and communication. Therefore, any modification in the serial code is immediately reflected in the parallel code.

After the local tree structure is constructed, a post-order traversal is performed on the tree structure and Particle-to-Multipole (\sptom\/) and Multipole-to-Multipole (\smtom\/) kernels are executed bottom up. The \sptom\ kernel is executed only at the leaf cells. It loops over all particles in the leaf cell to form the multipole expansion at the center of the leaf cell. The \smtom\/ kernel is executed only for the non-leaf cells. It loops over all child cells and translates the multipole expansions from its children's centers to its center.

Once the multipole expansions for all local cells have been determined, the multipole expansions are sent to the necessary processes in a sender-initiated fashion \cite{dubinski1996}. This reduces the latency by communicating only once, rather than sending a request to remote processes and then receiving the data. Such sender-initiated communication schemes were common in cosmological $N$-body codes since they tend to use only monopoles, and in this case the integer to store the requests is as large as the data itself if they were to use a request-based scheme. This data is used to construct the local essential tree (LET), that is, the union of all trees representing the entire domain as seen by the local process \cite{warren1992}. It gets coarser depending on the distance of the remote cell. In the present method, it is formed by simply grafting the root nodes of the remote trees. In conventional parallel FMM codes, a global octree is formed and partitioned using either HOT or ORB. Therefore, the tree structure was severed in many places, which caused the merging of the LET to become quite complicated. Typically, code for merging the LET would take a large portion of a parallel FMM code, and this made it difficult to implement new features such as periodic boundary conditions, mutual interaction, more efficient translation stencils, and dual tree traversals. \sexafmm\ is able to incorporate all these extended features and still maintain a fast pace of development because of this simplification in how the global tree structure is geometrically separated from the local tree structure.

While the remote information for the LET is being transferred, the local tree can be traversed. Conventional fast $N$-body methods overlap the entire LET communication with the entire local tree traversal. The LET communication becomes a bulk-synchronous \texttt{MPI\_alltoallv} type communication, where processes corresponding to geometrically far partitions send logarithmically less information, thus resulting in $O(\log{P})$ communication complexity where $P$ is the number of processes. Nonetheless, in traditional fast $N$-body codes this part is performed in a bulk-synchronous manner.

\section{Communication Reduction for the Adaptive Tree}\label{sec:communication_methodologies}
In the following sections, we present different novel techniques that can be used to do the hierarchically sparse data exchange ($\mathcal{HSDX}$) of the adaptive \sfmm\ tree, which are generally applicable to a variety of algorithms constituting definition~\ref{definition:hsde}. The optimization of global tree communication is essential to achieve strong scaling especially at a large scale. Such class of communication becomes very challenging due to the fact that \sexafmm\ has a highly optimized serial code that utilizes many-core parallelism, making the complete overlap with computation infeasible. The natural solution to this problem is to strong scale communication, but to our knowledge, it is not straightforward to achieve that for practical reasons such as network congestion, growing interaction lists, and the different implementations of some \smpi\ collectives that do not scale by definition e.g., \smpialltoallv\ . Therefore, it is important to look at these caveats while implementing a domain-specific communication scheme of the global \sfmm\ tree.

\theoremstyle{definition}\label{definition:hsde}
\begin{definition}
Let $\mathcal{T}$ be a global adaptive tree with $\mathcal{L}$ levels numbered from $l_0 - l_k$ (coarse to fine) and partitioned to $\mathcal{P}$ processes. $s$ is the “essential” subtree size such that $0<s<\mathcal{S}$. $P_i,P_j \subset{l_k}$, if the finest level $P_i,P_j$ share is k. We have a hierarchically sparse data exchange $\mathcal{HSDX}$ if for $P_i,P_j \subset{l_1}$ and $P_i,P_v \subset{l_2}$, $s_1<s_2$ and $s_1 != 0$
\end{definition}

\subsection{Overlapping Computation Depending on Communication Granularity} \label{sec:comm_grain_sec}
The importance of asynchronous communication arises from the fact that the latter is a limiting factor to performance at exascale especially when done collectively. This appears to be the case for hierarchical algorithms such as \sfmm\ and Multigrid method (MG). Hence, communication needs to be balanced and efficiently overlapped with local work. In \sfmm\ , it is known that a substantial amount of time is spent in doing local Multipole-to-Local (\smtol\/) and Particle-to-Particle (\sptop\/) computations, but the question is how often we need to communicate to reduce blocking for data given the problem size, distribution and scale. To answer this question, we have parametrized our \sfmm\ to accept different granularities of communication represented by the size of the LET's subset. The subsets may contain non-leaf cells requiring $O(p)$ steps for $p$ = order of multipole expansion (higher $p$ increases arithmetic intensity for low-level kernels) or leaf cells requiring $O(N/P)^{2}$ steps for $P$ = number of processes. The typical case would be to call a blocking \smpirecv\ on the expected tag because there is no useful work to do in the current context; however, since \smpi\ does not provide guarantees on the order of messages when used in mixed mode, our code consumes the available subtree and marks it as ``traversed". This mechanism will maximize concurrency and minimize the message queuing time. The calling task will keep traversing until requested cell is received or traversed by another task. 

Conventional parallel $N$-body methods use a bulk-synchronous \texttt{MPI\_alltoallv} to communicate the whole LET at once, and overlap this communication with the local tree traversal to hide latency. One could over-decompose the LET down to a per cell request, and then aggregate the communication to the optimal granularity. The bulk-synchronous communication model can be thought of as an extreme case of aggregation, while something like an RDMA per task per cell would be at the other end of the granularity spectrum.
There is a caveat: We still require further tuning to reduce global communication by indirectly relaying multipoles through neighbor processes as we will show in Sec.~\ref{sec:neighborhoodCollectiveSec} using Algorithm~\ref{multilevelAlg}.

\begin{figure*}[t] 
     \centering
     \subfloat[Hypercube communication of 256 MPI processes. Interactions get coarser as we move right and finish in LogP steps.]
     {\includegraphics[height=5cm,width=0.48\textwidth]{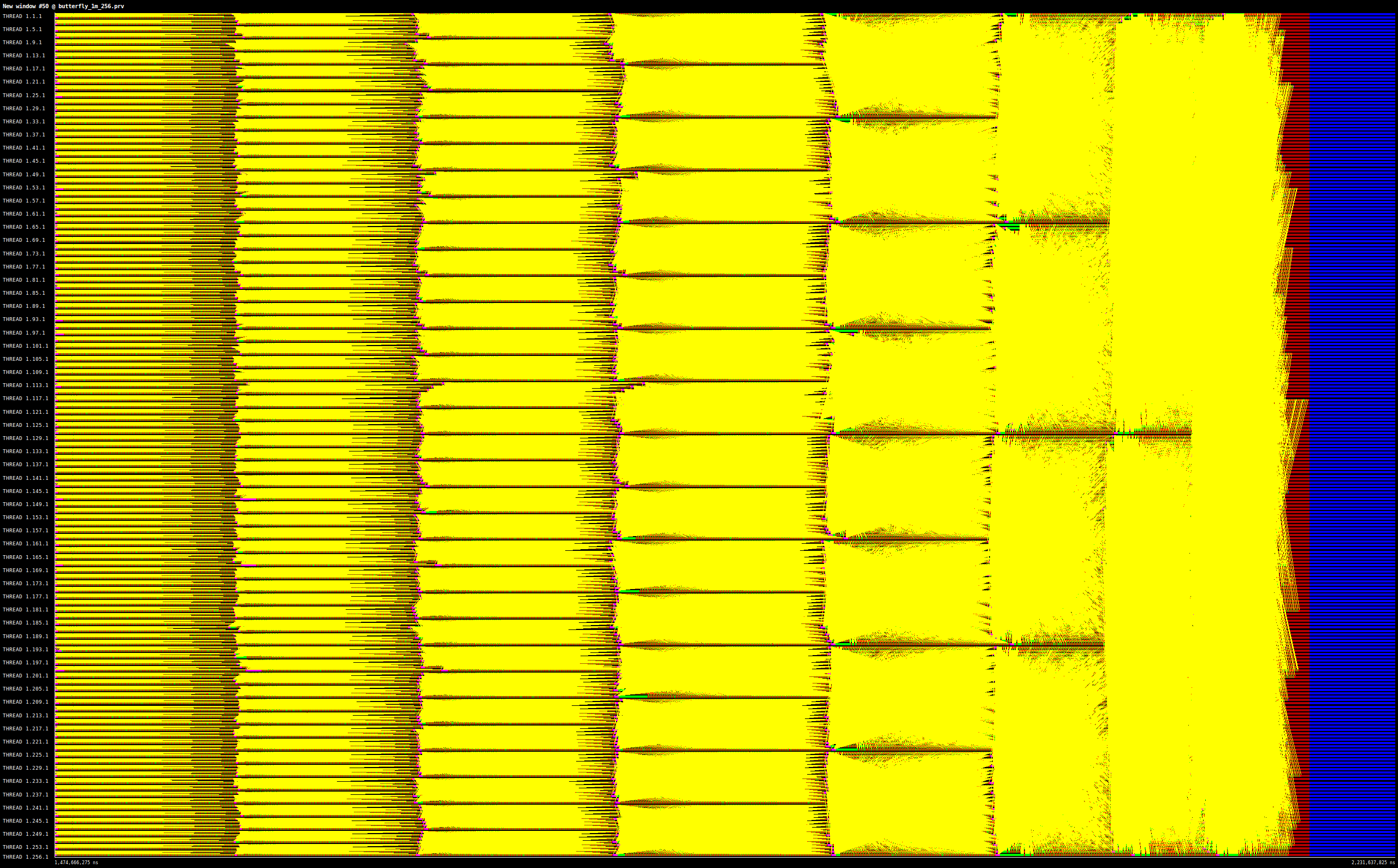}\label{fig:butterfly}}
     \hfill
     \subfloat[A zoomed-in version of the flat communication of 256 MPI processes, potentially causing contention when scaling on a supercomputer's network.]
     {\includegraphics[height=5cm,width=0.48\textwidth]{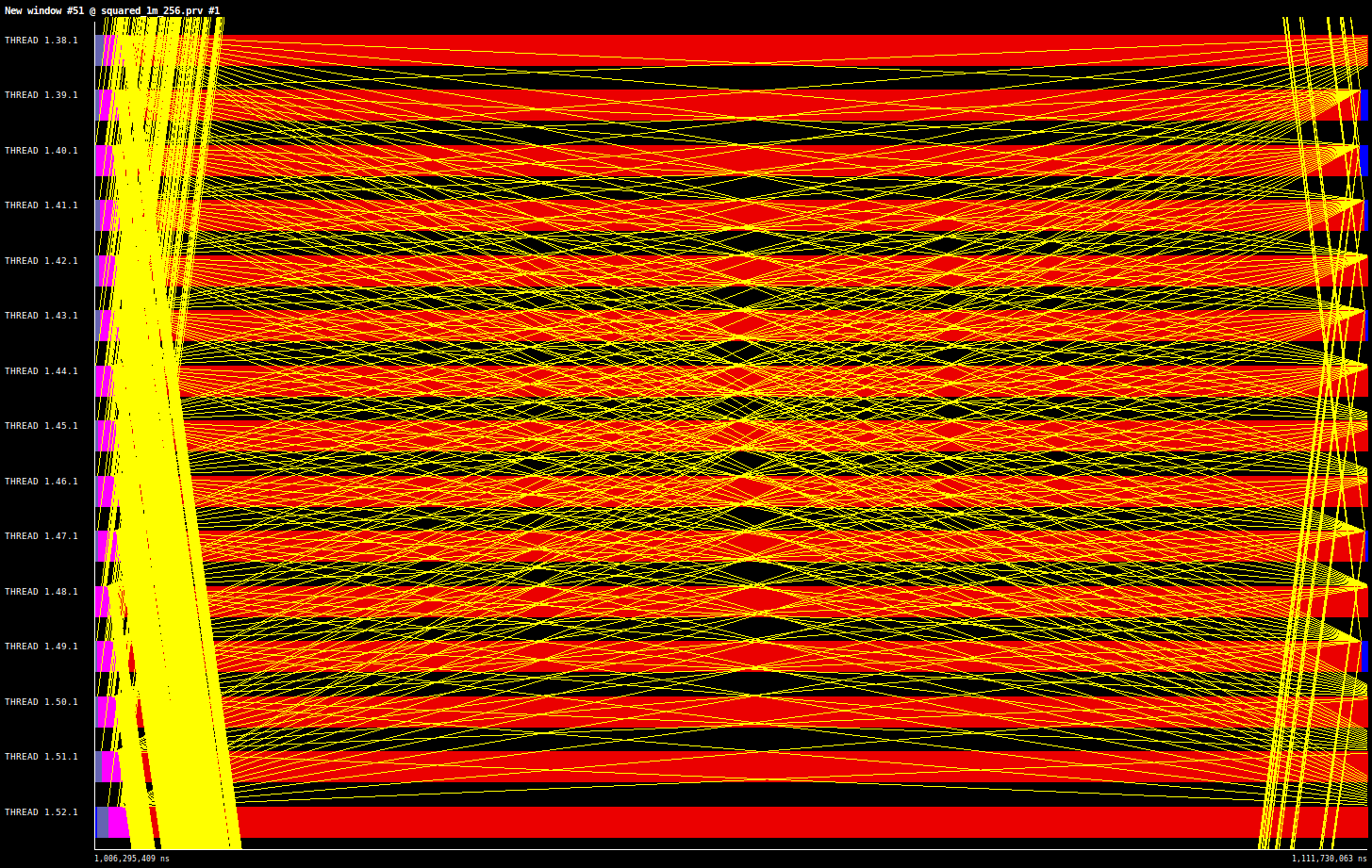}\label{fig:squared}}
\caption{Implemented communication patterns in \sfmm\ as visualized by Extrae .}
\label{fig:fmm_butterfly}
\end{figure*}

\subsection{Hierarchical Sparse Data Exchange Protocol ($\mathcal{HSDX}$)}\label{sec:neighborhoodCollectiveSec}

\begin{figure}[t]
\centering
\includegraphics[width=0.5\columnwidth]{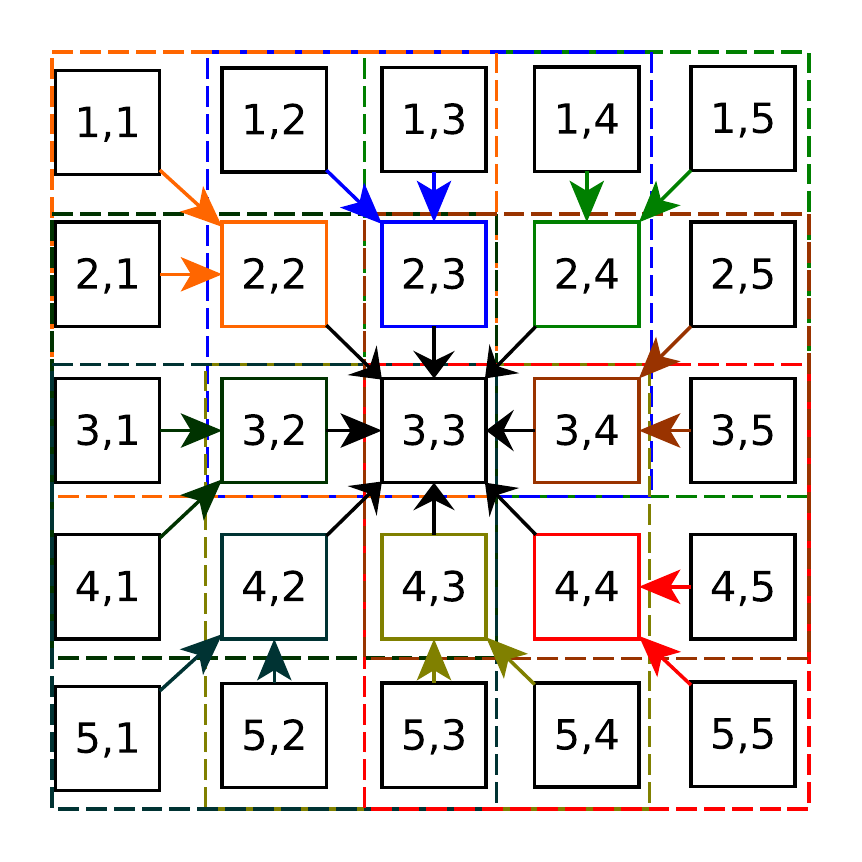}
\caption{The underlying data exchange graph of central process (3,3) within a uniform 2D grid of processes.}
\label{fig:neighborhood_collective}
\end{figure}

Lashuk et al. \cite{lashuk2009} define a set of parameters that denote the interaction lists, i.e., $U$-, $V$-, $W$- and $X$-lists of the \sfmm\ tree. The same analogy can be used for describing the relationship between adjacent processes such that exchanging the entire LET can happen in a few steps. The U- and V- lists constitute the adjacent nodes/processes through which global cells that contribute to the local tree are relayed. For the majority of the spatial $N$-Body partitioning methods, we can use the subdomain's bounding box to depict partitions that share a face, an edge or a vertex in $O(1)$ steps using Lemma~\ref{adjacancy_list_lemma}.
This enables us to create a breadth-first data exchange graph that starts from the local tree and covers all the cells from the essential tree.  Each node in the graph contains the corresponding partition id and the adjacent partition id, which is needed since communication strictly happens between adjacent nodes. Figure~\ref{fig:neighborhood_collective} shows the exchanges needed to receive the entire LET by target process (3,3), with overlapping direct clusters enclosed in dashed squares. The corresponding data exchange graph of node (3,3) contains a node with id (1,5) and an adjacent id of (2,4), meaning that cell data of (1,5) can be acquired through (2,4) in the second stage of exchanges. To inherently achieve algorithmic balance, we hardwire edges in such a way that messages are evenly distributed over direct neighbors. If we start with direct neighbor (2,4), a naive approach would exhaust all its direct neighbors, namely \{(1,3),(1,4),(1,5),(2,5),(3,5)\}, thus overloads its buffers and causes imbalance. The next neighbor (3,4) will only have (4,5) data to relay. Therefore, we design our communication graph such that for internal processes in a uniform domain, the average number of messages received from direct neighbors in each step is $\ceil[\Big]{\frac{5^{D}-3^{D}}{3^{D}-1}}$. Using notations from Table~\ref{multilevelAlgSymbols}, we can generalize this formula to non-uniform domains if we turn it into 

\begin{equation}\label{equation:neighbor_count}
NB=\ceil[\Big]{\frac{\tau(P,1) - \zeta(\Omega(P))}{\zeta(\Omega(P))-1}}
\end{equation}

We finally reach a stage where each process has access to the near and far-field interactions, thus accomplishing global communication using multiple calls to \smpialltoallvn. Algorithm~\ref{multilevelAlg} and Table~\ref{multilevelAlgSymbols} summarize our method.

\subsubsection{Time Complexity of the adaptive $\mathcal{HSDX}$}
A good lower bound complexity for $\mathcal{HSDX}$ is $\mathcal{NBX}$ i.e. $\Omega(\log{P})$ from \cite{hoefler2010}, when non-neighbor data exchange is extremely sparse or non-existent. The hierarchical sparsity in Definition~\ref{definition:hsde} increases as we move away from target processes. The data exchange graph can be mapped to a tree since there is exactly one path from $P_i$ to $P_j$, with an order bounded by Eq~\ref{equation:neighbor_count}. An upper bound is analogous to a fully dense communication, such that $O(\log{P})$ exchanges happen $O(\log{P})$ times, which is equivalent to $O(\log^2{P})$. Table~\ref{tab:complexity} shows \sfmm\ communication complexity for uniform domains.

\begin{lemma} \label{adjacancy_list_lemma}
A partition $P'$ is added to the adjacency list of $P$ iff for any dimension $D$ $maxBound(P'_x)-maxBound(P_x)>\epsilon$ and $minBound(P_x)-minBound(P'_x)>\epsilon$
\end{lemma}

\begin{table}[t]
\caption{$\mathcal{HSDX}$ Algorithm Communication Symbols.}\label{multilevelAlgSymbols}
\begin{center}
\begin{tabular}{|c|c|}
\hline
Symbol & Indication\\
\hline \hline
$P$ and $P'$ & local and global partitions\\
\hline
$\Omega(P)$ & subdomain boundary\\
\hline
$\zeta(\Omega(P))$ & direct neighbors of P\\
\hline
$\mathcal{T}$ & level-by-level communication adjacency graph\\
\hline
\end{tabular}
\end{center}
\end{table}

\begin{algorithm} 
\SetKwFunction{BuildCommTree}{BuildCommTree}
\SetKwInOut{Input}{input}
\SetKwInOut{Output}{output}
\caption{$\mathcal{HSDX}$ - Hierarchical Sparse Data Exchange} \label{multilevelAlg}
\Input{A list $l_{in}$ of cells and destinations}
\Output{A list $l_{out}$ of cells and sources}
\BlankLine
\ForEach{$P'$ in $\Omega(P,\beta)$}{
  add(P',$\zeta(\Omega(P))$)
}
$\mathcal{T} \leftarrow$ \BuildCommTree{$\zeta(\Omega(P))$}\;
create distributed MPI graph topology\;
\ForEach{$l$ in $\mathcal{T}.Levels$}{
  \ForEach{$P'$ in $\zeta(\Omega(P))$}{
    reduce tree based on the bounding box and forward to P'\;
  }
  exchange meta data\;
  call \smpialltoallvn \;
}
\end{algorithm}

\subsection{Pairwise Exchange for Reducing Contention}\label{sec:contention}
It is observed at large scale that direct communication between sources and targets results in network contention which can be amortized by relaying multipoles through neighbor processes while utilizing the well-known pattern of $N$-Body interactions. Therefore, to mimic $O(\log{P})$ complexity for boundary distributions, we implement a modified version of the well-known hypercube (butterfly) global communication scheme which starts out by the fine neighbor interactions depicted by ($P \xor 2^i$) and gets coarser as we move towards the $\mathcal(\log P)$ step. This is clearly visualized in Fig~\ref{fig:fmm_butterfly} using Extrae, a tool that uses different interposition mechanisms to inject probes into the target application so as to gather information regarding the application performance. During this work, the tool is used to better understand the performance of the application pertaining to the used communication techniques. In Fig~\ref{fig:fmm_butterfly} , the horizontal axis represents the visualized timeline and the vertical axis represents the MPI processes. The yellow colors declare communication links, whereas the blue colors represent computation and red color symbolizes MPI\_Wait calls. One of the main advantages of carrying out communication in $\mathcal(\log P)$ steps, as in is Fig~\ref{fig:butterfly}, is that subtrees received at intermediate stages can be asynchronously traversed, which otherwise cannot be done if communication is done with blocking collectives. 

\begin{table}[b]
\caption{Communication complexity of FMM.}
\label{tab:complexity}
\begin{center}
\begin{tabular}{|c|c|}
\hline
Reference & Communication complexity\\
\hline \hline
Teng (1998) \cite{teng1998} & $\mathcal{O}\left(P(N/P)^{2/3}(\log N+\mu)^{1/3}\right)$\\
\hline
Lashuk \textit{et al.} (2009) \cite{lashuk2009} & $\mathcal{O}\left(\sqrt{P}(N/P)^{2/3}\right)$\\
\hline
Lashuk \textit{et al.} (2012) \cite{lashuk2012} & $\mathcal{O}\left(\log{P}+(N/P)\right)$\\
\hline
Yokota \textit{et al.} (2014) \cite{yokota2014} & $\mathcal{O}\left(\log P+(N/P)^{2/3}\right)$ \\
\hline
\end{tabular}
\end{center}
\end{table}

\section{Performance Analysis}
\subsection{Experimental Setup}
Our experiments are on Shaheen XC40, the rank 15 supercomputer according to the November 2016 Top500 list, located at King Abdullah University of Science and Technology. It has 196,608 physical cores and HPL performance of 5.537 PFlop/s. Each node is equipped with dual socket Intel Xeon E5-2698v3 16C 2.3GHz and Cray Aries interconnect with dragonfly topology. \\

Throughout the following experiments, the underlying \sfmm\ code is compiled with the Laplace kernel, Cartesian coordinates, $P$ = 4 (order of expansion) and spherical boundary distribution unless otherwise stated. Problems have been partitioned using the hybrid partitioning from section \ref{sec:hybrid}. To demonstrate the effectiveness of the presented methodologies, we start by showing how optimal grain size for a specific problem is chosen, then assessing the scalability with the tuned granularity of communication. Then, results from using $\mathcal{HSDX}$ vs. existing communication reducing approaches are presented. Good scalability shows that an inordinate cost is not paid for intra-node communication, as opposed to the conventional bulk-synchronous approach, for which performance depends on the underlying network topology, the implementation of collectives like \salltoall\ or \sallgather\/, the available memory size and bandwidth, and the frequency at which synchronization is triggered. 

\begin{figure}[t]
	\centering
	\includegraphics[width=0.7\textwidth]{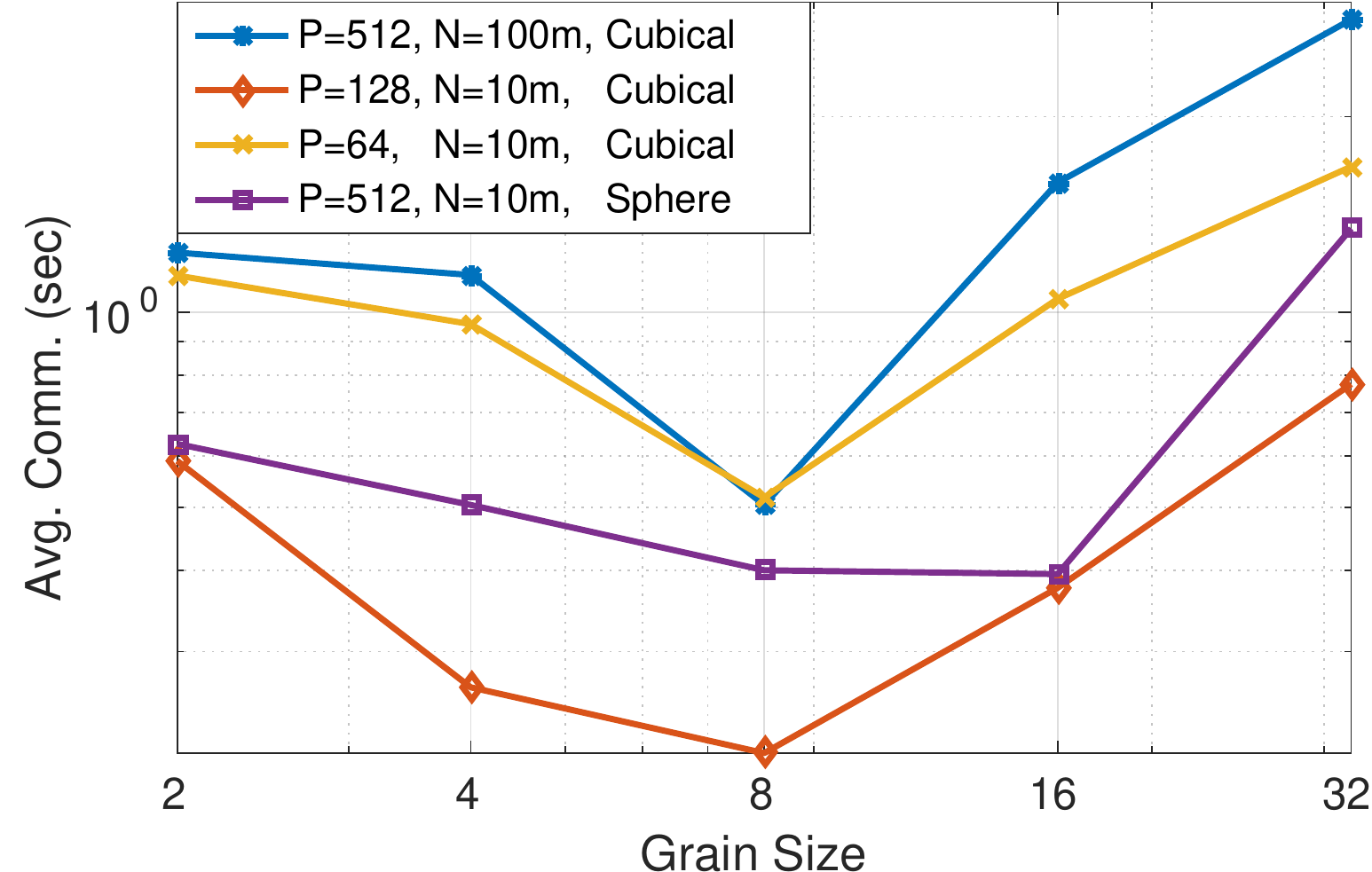}
	\caption{Average communication time for different sizes and distributions as grain-size is varied.}
	\label{fig:optimal_grain}
\end{figure}

\begin{figure}[t]
	\centering     
	\includegraphics[width=0.7\textwidth]{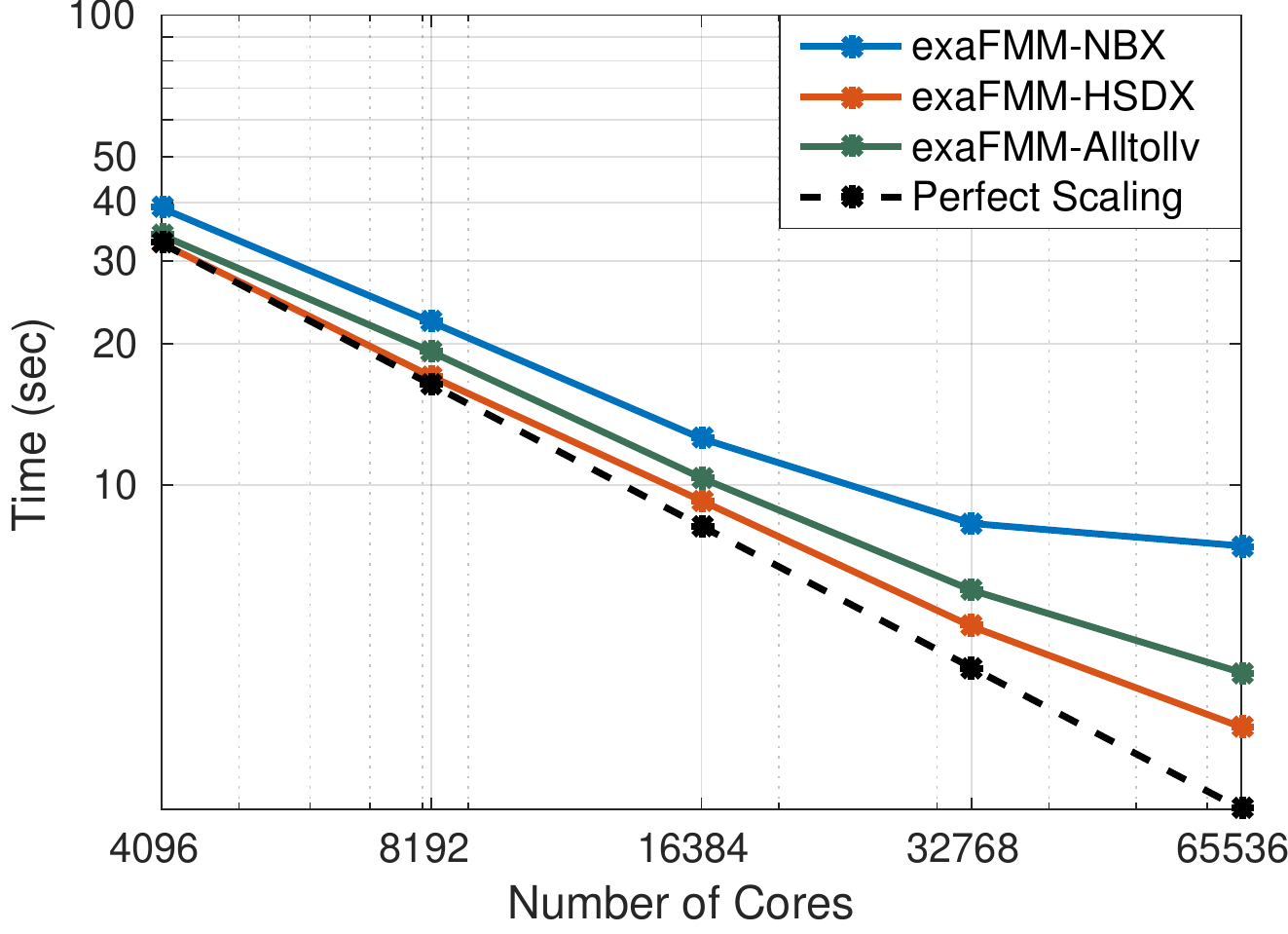}
	\caption{Strong scaling across different communication protocols with $10^{10}$ particles.}\label{fig:strong_scalability}
\end{figure}

\begin{table}[t]
	\caption{$\mathcal{HSDX}$ strong scalability analysis with \sfmm\ .}\label{table:scalability_analysis}
	\begin{center}
		\begin{tabular}{|c|c|c|c|c|c|}
			\hline
			$P$ & $4,096$ &$8,192$ & $16,384$ & $32,768$ & $65,536$ \\
			\hline \hline
			HSDX& $32.72$ & $17.02$ & $9.27$ & $5.008$ & $3.05$ \\
			\hline
			Rel. Speedup& $1$ & $1.92$ & $3.53$ & $6.53$ & $10.70$ \\
			\hline
			Effeciency&-& $0.96$ & $0.88$ & $0.81$ & $0.66$ \\
			\hline
			Enhancement& $3.87\%$ & $11.41\%$ & $10.55\%$ & $16.27\%$ & $23.44\%$ \\
			\hline
		\end{tabular}
	\end{center}
\end{table}
 
\subsection{Communication Time for Different Granularities}\label{sec:comm_grain}
In order to show the direct effect of asynchronous traversal on performance, we gradually vary the grain-size and measure the communication time, which is the most dominant factor at a large scale. Optimal granularity is a tuning parameter that varies with problem size, distribution and other factors as depicted by the average communication time in Fig.~\ref{fig:optimal_grain}, where subtree size is gradually increased. The theoretical maximum size is the entire LET. We stop at a certain threshold (32 in this case) because when it is increased further, a huge jump in time occurs. This is attributable to the change in communication protocol as per the Cray\textsuperscript{\textregistered} MPICH specification from Eager Message to Rendezvous Message Protocol. When the message size exceeds a specific threshold (8 KB in this case), MPICH2 GNI NetMod alters the pathways towards a more relaxed algorithm for point-to-point inter-node messaging. A similar approach is developed in other \smpi\ implementations like Open MPI and Intel\textsuperscript{\textregistered} MPI. Hence, the remote tree traversal enables us to tune the performance by reducing the communication time enough to increase the impact of latency hiding.

\subsection{Scalability of Spherical Boundary Distribution with $\mathcal{HSDX}$ }\label{sec:scalability_fine_grain}
In Fig.~\ref{fig:strong_scalability}, we test the strong scalability at optimal grain size using $\mathcal{HSDX}$ for a large problem of $10^{10}$ particles. It follows that we have an efficient asynchronous communication when remote calls are non-blocking, have tunable granularity and when control is handed over to useful work rather than waiting immediately. To show this, we have integrated and compared several communication protocols within \sexafmm\ in Fig.~\ref{fig:strong_scalability}. We note that $\mathcal{HSDX}$ is the closest to ideal scaling and has the advantage of fastest time-to-solution since it limits the inter-rack communication penalty on the dragonfly network by solely exchanging data through neighbors. By just looking at Fig.~\ref{fig:strong_scalability}, it is hard to see that $\mathcal{HSDX}$ is at potential advantage for the exascale era. So we find it crucial to present Table~\ref{table:scalability_analysis} that shows a more detailed analysis of the strong scalability. We notice a 6-fold increase in performance gain (from $3.87\%$ to $23.44\%$) over the corresponding \smpialltoallv\ implementation as more cores are added. The parallel efficiency decreases, however, as the problem gets smaller while communication overhead prevails. Conventional $O(P)$ communication schemes stop scaling after 2048 nodes (65,536 cores) of Shaheen XC40. According to our largest setup that has an input of $N=10^{10}$, we have an update rate of approximately $10^{9}$ particles/second. 

\begin{figure}[t]
	\centering
	\subfloat[Test][Big example comparison of different communication protocols when weak-scaling 15m particles.]
	{\includegraphics[width=0.45\textwidth]{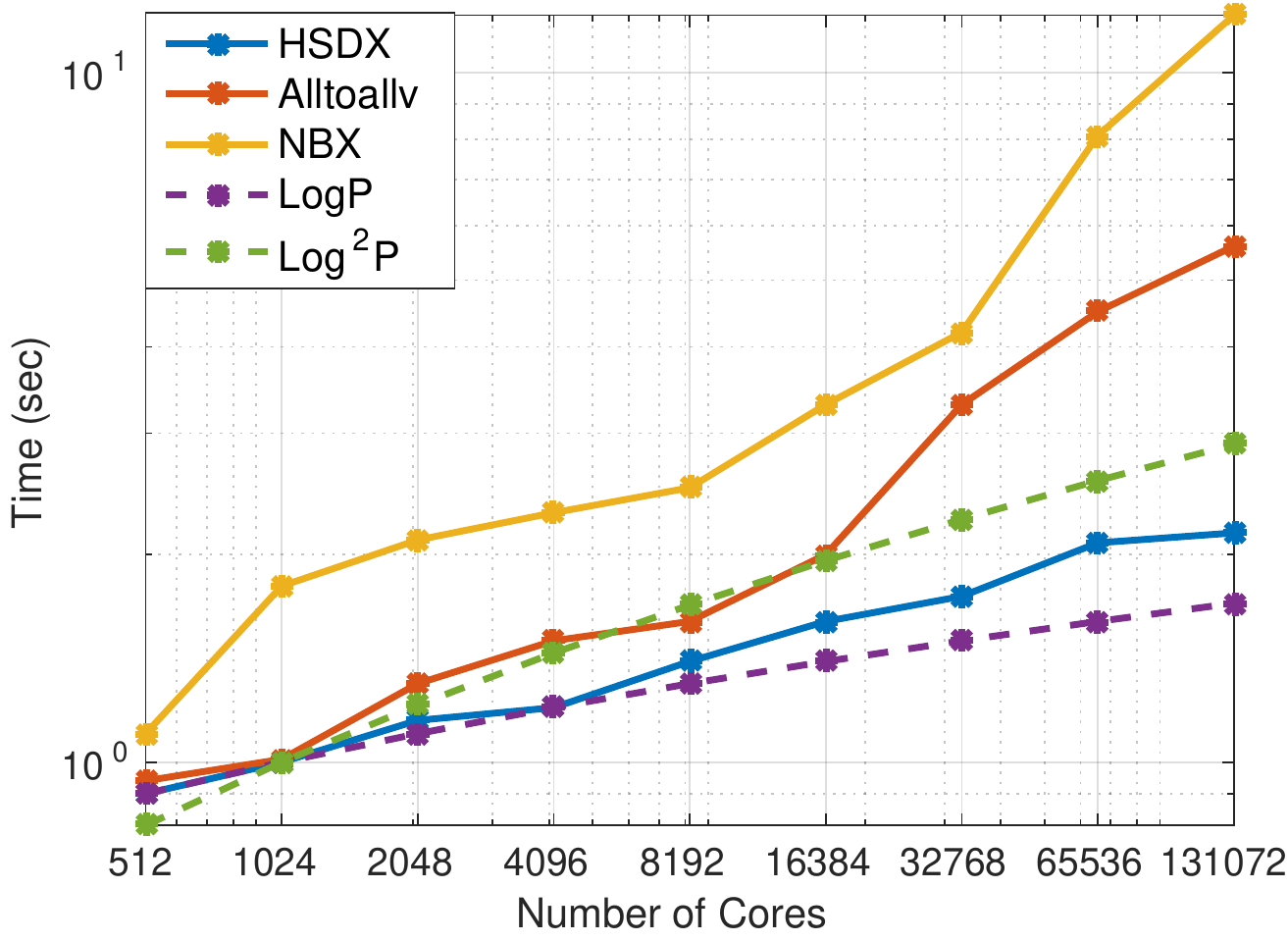}\label{fig:alltoallv_vs_neighborhood1}} \hfill
	\subfloat[Test][Small example comparison of different communication protocols when weak-scaling 200k particles]
	{\includegraphics[width=0.45\textwidth]{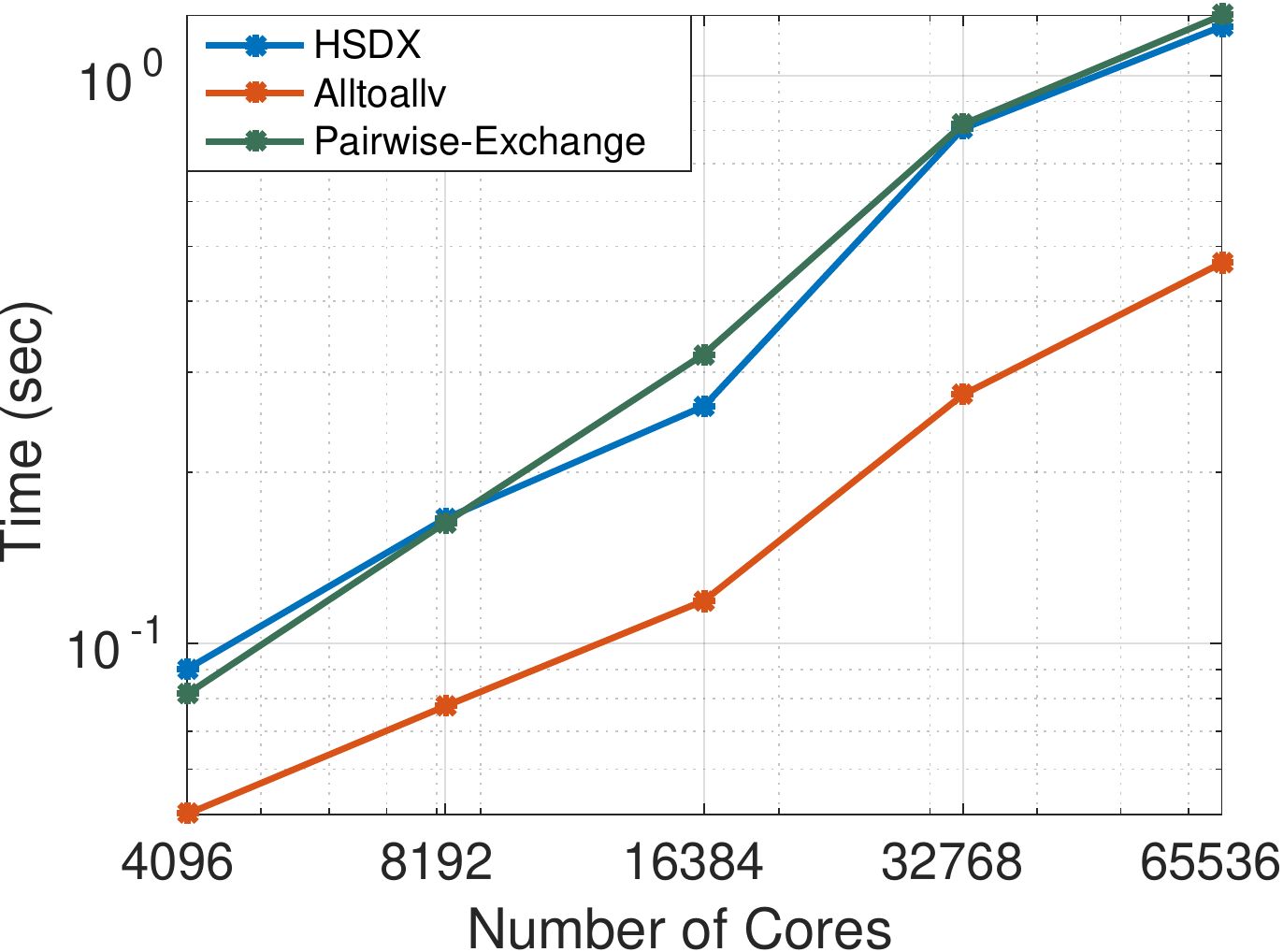}\label{fig:alltoallv_vs_neighborhood2}}
	\caption{Communication scaling for big and small examples.}
\end{figure}

\begin{figure*}[t]
\centering
\subfloat[Test][\sexafmm\ vs \spvfmm\ breakdown.]
{\includegraphics[width=0.35\textwidth,height=3.5cm]{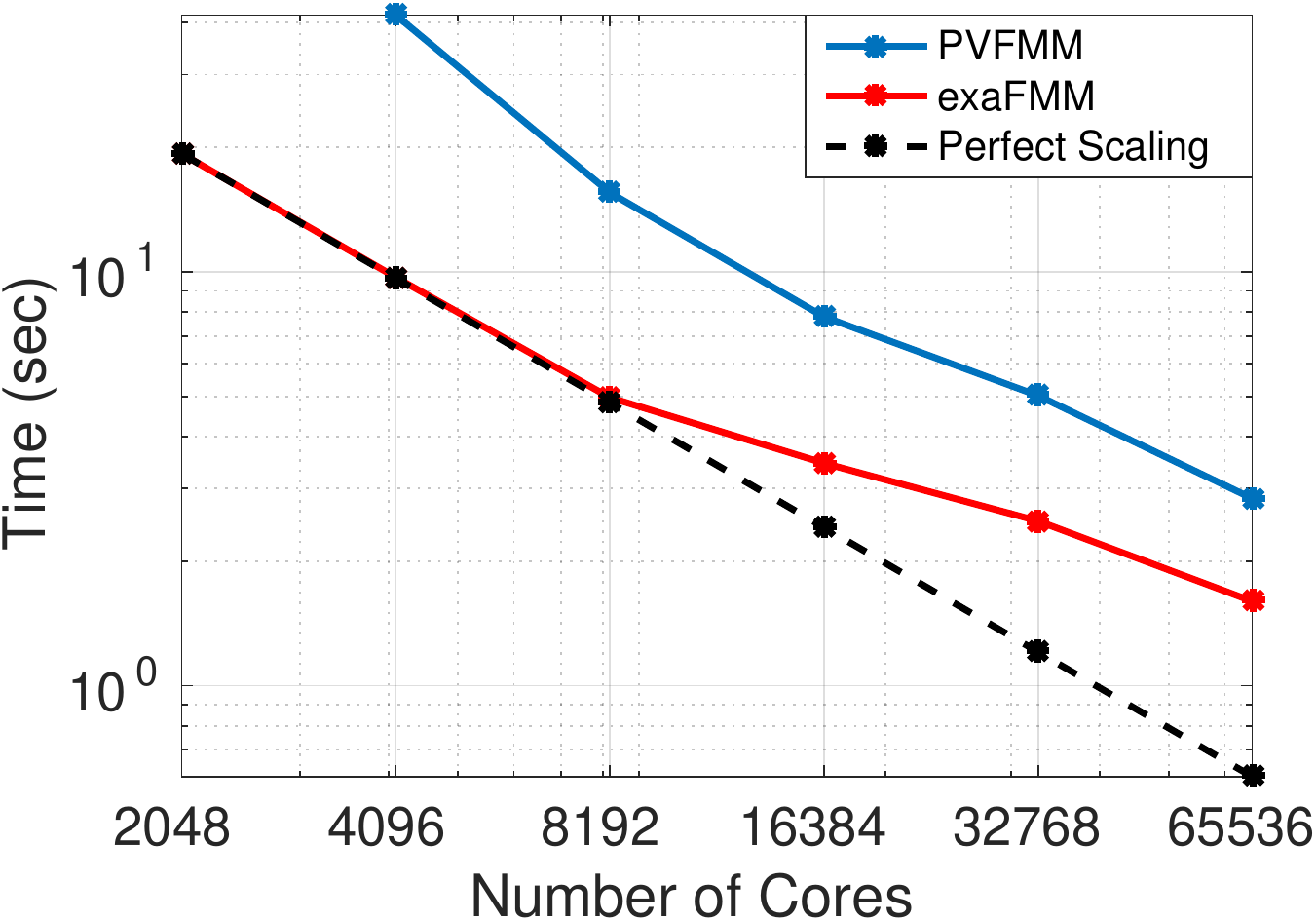}\label{fig:exafmmvspvfmm}}
\subfloat[Test][\spvfmm\ breakdown.]
{\includegraphics[width=0.35\textwidth,height=3.5cm]{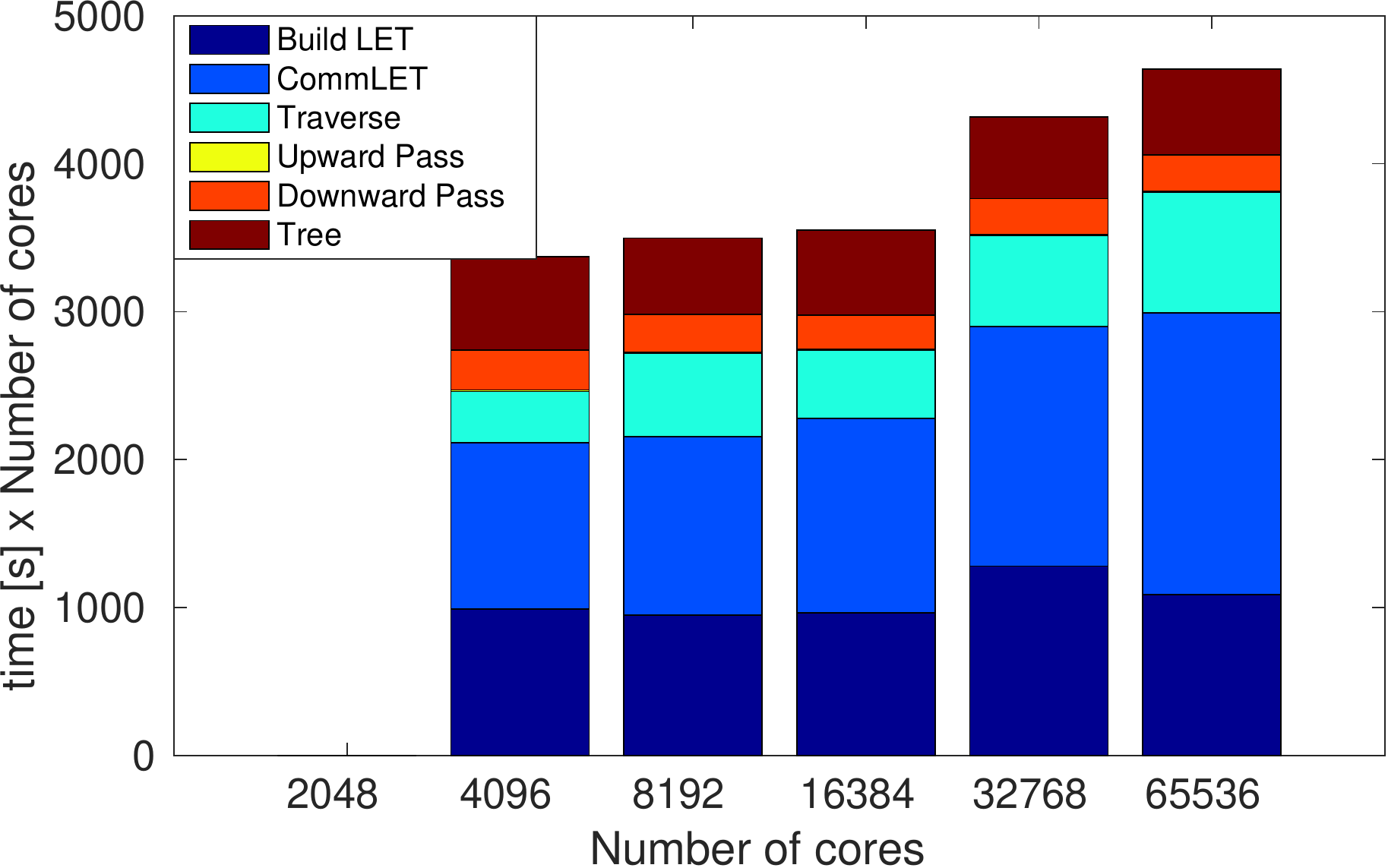}\label{fig:pvfmm_breakdown}}
\subfloat[Test][\sexafmm\ breakdown.]
{\includegraphics[width=0.35\textwidth,height=3.5cm]{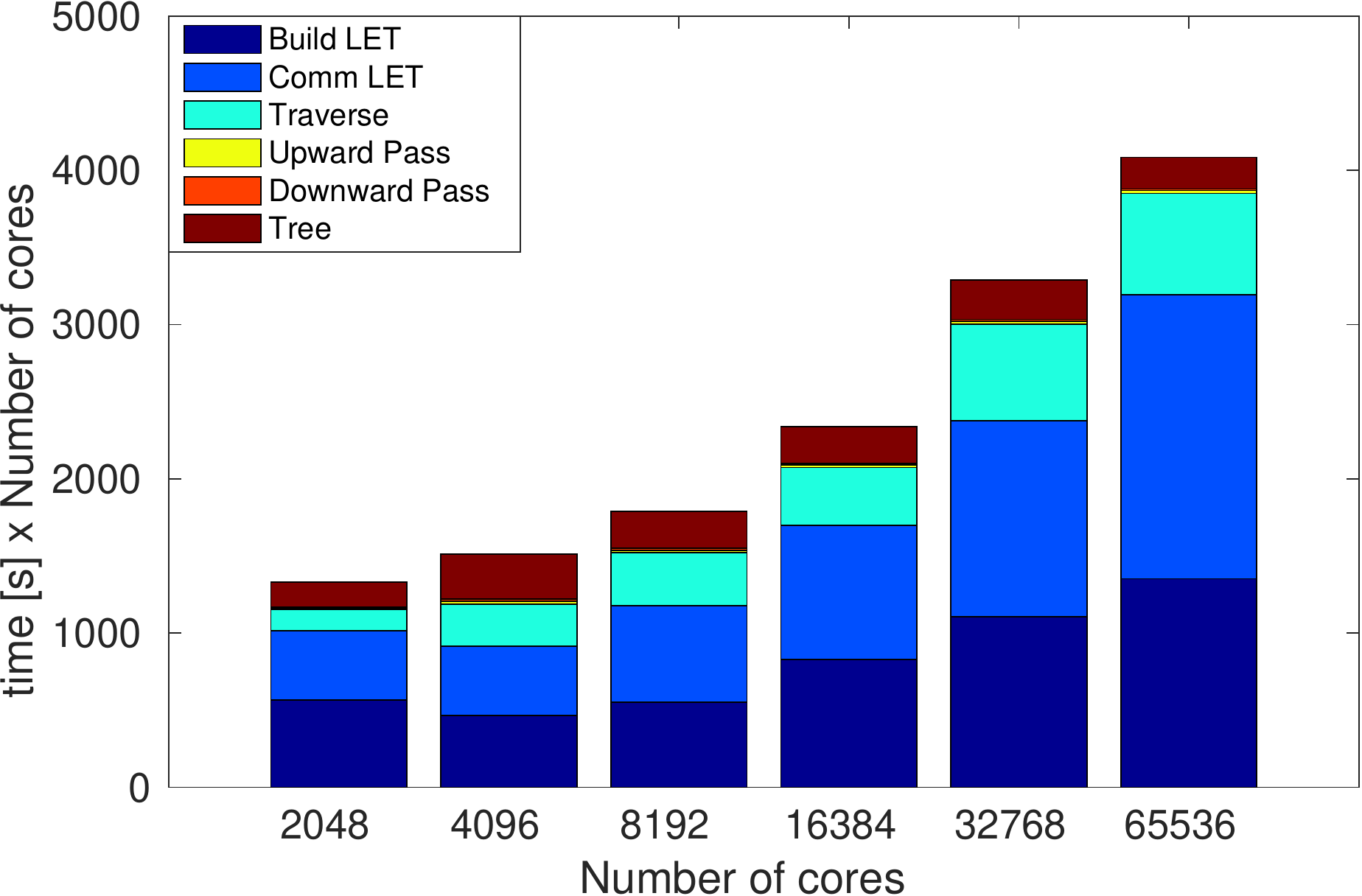}\label{fig:exafmm_breakdown}}
\caption{Strong scaling $2^{31}$ unknowns for sphere distribution and $P=4$ and comparing \sexafmm\ while using $\mathcal{HSDX}$ communication and \spvfmm\ .}
\label{fig:exafmmvspvfmm_breakdown}
\end{figure*}

\spvfmm\ is a large-scale FMM library that uses a kernel independent implementation, thus widens its target range of applications that require calculation of potential for elliptic kernels \cite{malhotra2015pvfmm}. In this experiment, we attempt to compare the strong-scaling performance of \spvfmm\ to our \sexafmm\ branch. It is worth noting that citing independent work is not meant to deem one superior to the other, but on the contrary, it is to give rise to our promising performance boosting strategies that tackle problematic communication and partitioning issues that are likely to arise in the near exascale era. In their most recent reports on \spvfmm\/, Malhotra et al. \cite{malhotra2015pvfmm} report perfect scalability up to 256 cores when running the Laplace kernel to compute potentials for $10^{8}$ distributed on the surface of an ellipsoid. From that point onwards, communication cost starts to grow. They achieve 95\% speedup corresponding too about 37\% parallel efficiency.
We switch to neighborhood collective communication presented in Section~\ref{sec:neighborhoodCollectiveSec} for this comparison, since it vastly reduces network contention by propagating cells through direct neighbors only. Fig.~\ref{fig:alltoallv_vs_neighborhood1} shows consistent weak-scalability of communication over the conventional \smpialltoallv\ implementation. The presented approach shows a faster time-to-solution in \sexafmm\ vs. \spvfmm\ when computing 2 billion unknowns as in Fig.~\ref{fig:exafmmvspvfmm_breakdown}. We cannot claim that scaling will persist indefinitely beyond the depicted number of cores, but when we have an exascale application that requires orders of magnitude larger problems that can fit in the machine's memory, we have a strong evidence of strong scalability.

\subsection{Evaluation of Neighborhood Collective Communication using $\mathcal{HSDX}$}\label{sec:scalability_neighborhood}
Fig.~\ref{fig:alltoallv_vs_neighborhood1} compares $\mathcal{HSDX}$ using neighborhood collectives to $\mathcal{NBX}$ and \smpialltoallv\/. For the class of problems that constitute a hierarchically sparse data exchange defined in \ref{definition:hsde}, $\mathcal{HSDX}$ is asymptotically bounded by the $c_{1}\log{P}$ and $c_{2}\log^{2}{P}$. This behavior is shown for the boundary distribution solving Laplace Cartesian \sfmm\ kernels with $P=4$ (order of expansion). However, the figure does not suggest that $\mathcal{HSDX}$ can generally replace its rivals; we still believe that $\mathcal{NBX}$ would outperform our algorithm in the general sparse data exchange, because it has the advantage of both $O(\log{P})$ upper bound in addition to the use of a non-blocking barrier and synchronized sends \cite{hoefler2010}.

Fig.~\ref{fig:alltoallv_vs_neighborhood2} weak scales a small example in order to reduce the effect of non-neighbor communication. The fact that $\mathcal{HSDX}$ and Pairwise exchange exhibit similar performance is anticipated since they almost have identical $logP$ behavior in such cases. They seem to lose herein against \smpialltoallv\/ because of the initialization overhead included in communication time.

\section{Conclusion}
In this work, we propose algorithms that improve data locality, remote data access, and load-balance of the $N$-body problem. These algorithms contribute to producing an \sfmm\ solver that exploits communication redundancy and computation overlap. We show that Hilbert space-filling curves may not be the most optimal choice to partition boundary domain distributions. $\mathcal{HSDX}$ shows good strong and weak scalability for large adaptive hierarchically sparse problems, and falls within proven asymptotic time complexities. Shared memory parallelism is important to utilize resources within a node and to alleviate the problems with \smpi\ resource management; thus we need to consider it in future implementations. We are working on improving  $\mathcal{HSDX}$ so that it exploits the advantages of $\mathcal{NBX}$ to widen its range of use cases. As for application, we are intending to make the presented solver a part of an \sfmm\ preconditioner for the Poisson equation, which has variety of applications in diffusive and equilibrium processes in fluid dynamics and many other applications. 

\bibliographystyle{IEEEtran}
\bibliography{references}

\end{document}